\documentclass[useAMS,usenatbib]{mn2e}
\usepackage{graphicx}

\title[On the oldest asteroid families in the main belt]
{On the oldest asteroid families in the main belt}
\author[V. Carruba, D. Nesvorn\'{y}, S. Aljbaae, R. C. Domingos, M. Huaman]{V. Carruba$^{1,3}$\thanks{E-mail: vcarruba@feg.unesp.br}, D. Nesvorn\'{y},$^{3}$, S. Aljbaae$^{1}$, R. C. Domingos$^{2,1}$, M. Huaman$^{1}$\\
$^{1}$UNESP, Univ. Estadual Paulista, Grupo de din\^{a}mica Orbital e
  Planetologia, Guaratinguet\'{a}, SP, 12516-410, Brazil. \\
$^{2}$UNESP, Univ. Estadual Paulista, S\~{a}o Jo\~{a}o da Boa Vista, SP, 
13874-149, Brazil.\\
$^{3}$Department of Space Studies, Southwest Research Institute, Boulder, 
CO, 80302, USA.\\
}

\begin{document}

\date{Accepted 2016 March 2.  Received 2016 March 2; in original form 2015 December 3.}

\pagerange{\pageref{firstpage}--\pageref{lastpage}} \pubyear{2015}

\maketitle

\label{firstpage}

\begin{abstract}
Asteroid families are groups of minor bodies produced by 
high-velocity collisions.  After
the initial dispersions of the parent bodies fragments, their orbits
evolve because of several gravitational and non-gravitational effects,
such as diffusion in mean-motion resonances, Yarkovsky and YORP effects, 
close encounters of collisions, etc. The subsequent dynamical 
evolution of asteroid family members may cause some of the original fragments 
to travel beyond the conventional limits of the asteroid family.  
Eventually, the whole family will dynamically disperse and no longer be 
recognizable.  

A natural question that may arise concerns the timescales for dispersion of
large families.  In particular, what is the oldest still recognizable family 
in the main belt?  Are there any families that may date from the late stages
of the Late Heavy Bombardment and that could provide clues on our understanding
of the primitive Solar System?  In this work, we investigate the dynamical 
stability of seven of the allegedly oldest families in the asteroid main 
belt.  Our results show that none of the seven studied families has a 
nominally mean estimated age older than 2.7 Gyr, assuming
standard values for the parameters describing the strength of the Yarkovsky 
force.  Most ``paleo-families'' that formed between 2.7 and 3.8 Gyr would 
be characterized by a very shallow size-frequency distribution,
and could be recognizable only if located in a 
dynamically less active region (such as that of the Koronis family).  
V-type asteroids in the central main belt could be compatible with a 
formation from a paleo-Eunomia family.

\end{abstract}

%

\section{Introduction}
\label{sec: intro}

Asteroid families are born out of collisions.  They ``die'' when the fragments 
formed in the collisional event disperse so far because of gravitational
or non-gravitational forces that the family is no longer recognizable as a 
group.  The time needed to disperse small groups has been the subject
of several studies (from our group, see for instance Carruba et al. 2010a, 
2011, and 2015).  Less attention has been given to the dispersion of larger 
families.  Recent studies (Brasil et al. 2015) suggested that no family
could have likely survived the Late Heavy Bombardment (LHB hereafter, at 
least 3.8 Gyr ago), in the jumping Jupiter scenario, which set an upper
limit on the maximum possible age of any asteroid family.
In this work we focus our attention on the families whose estimated 
age, according to Bro\v{z} et al. (2013) could possibly date from 
just after the LHB, and whose existence, 
if confirmed, could provide precious clues on the early stages of our Solar 
System.   

Bro\v{z} et al. (2013) identified 12 families whose age estimate 
might have been compatible with an origin during the LHB, or just after:
the Maria, Eunomia, Koronis, Themis, Hygiea, Meliboea, Ursula, Fringilla,
Alauda, Sylvia, Camilla, and Hermione families.  The Sylvia family and the
proposed long-lost groups around Camilla and Hermione in the Cybele region
were recently studied in Carruba et al. (2015), and will not be treated in 
detail here.  That work found that, while all asteroids in the Cybele region
were most likely lost during the jumping Jupiter phase of the model of 
planetary migration of Nesvorn\'{y} et al. (2013), some of the largest fragments
($D > 5$~km, with D the body diameter) of a hypothetical post-LHB 
Sylvia family may have remained in the Cybele region, but the identification
of these groups could be difficult.  Due to local dynamics, any Camilla and 
Hermione families would disperse in time-scales of the order of 1.5 Gyr.  
The Fringilla family is a rather small group (134 members, according
to Nesvorn\'{y} et al. 2015) in the outer main belt.  
Since we are already studying the larger Themis, Meliboea, Alauda and 
Ursula families in the outer main belt,  and since the group is
not large enough for the techniques used in this work, we will not further 
investigate this family in this paper.  The Hygiea family was studied at 
length in Carruba et al. (2014a), that found a maximum possible age of 3.6 Gyr, 
just a bit younger than the minimum currently believed epoch of the last stages
of the LHB, i.e., 3.8 Gyr (Bottke et al. 2012; also the age of the Hygiea 
family should be most likely younger than 3.6 Gyr, because of the long-term 
effect of close encounters with 10 Hygiea and of the stochastic YORP
effect of Bottke et al. 2015, this last effect not considered in 
Carruba et al. 2014a).  Age estimates for the other families listed in 
Bro\v{z} et al. (2013) are however known with substantial uncertainty, and were
obtained before Bottke et al. (2015) modeled the so-called stochastic
YORP effect, in which the shapes of asteroids changes 
between YORP cycles, and whose effect is generally 
to reduce the estimates of the
given asteroid family age.  Understanding which, if any, of the proposed
``Bro\v{z} seven'' oldest groups (Maria, Eunomia, Koronis, Themis, 
Meliboea, Ursula, Alauda) could have survived since the LHB, and how,
is the main goal of this work.
 
This paper is so divided: In Sect.~\ref{sec: fam_ide} we identified
the seven ``Bro\v{z}'' families in a new space of proper elements
and SDSS-MOC4 $gri$ slope and $z' -i'$ colors, and use the method of 
Yarkovsky isolines to obtain estimates of maximum possible ages of these
groups.  In Sect.~\ref{sec: chron} we use Monte Carlo methods (see
Carruba et al. 2015) to obtain refined estimates of the family age and 
ejection velocity parameters of the same families.  
Sect.~\ref{sec: dyn_old} deals with the dynamical evolution and
dispersion times of fictitious simulated ``Bro\v{z}'' groups since
the latest phases of the Late Heavy Bombardment, when the 
stochastic YORP effect  (Bottke et al. 2015) and past changes in
the solar luminosity (Carruba et al. 2015) are accounted for. 
Finally, in the last section, we present our conclusions.

\section{Family identification}
\label{sec: fam_ide}

In order to obtain age estimates of the seven Bro\v{z} families, we first
need to obtain good memberships of these groups.  Any method used should
aim to reduce the number of possible interlopers, objects that may be
part of the dynamical family, but have taxonomical properties inconsistent
with that of the majority of the members.  DeMeo and Carry (2013) 
recently introduced a new classification method, 
based on the Bus-DeMeo taxonomic system, that employs Sloan Digital Sky 
Survey-Moving Object Catalog data, fourth release (SDSS-MOC4 
hereafter, Ivezi\'{c} et al. 2002) $gri$ slope 
and z' -i' colors.  In that article the authors used the photometric 
data obtained in the five filters $u', g', r', i'$, and $z'$, from 0.3 to 1.0 
$\mu m$, to obtain values of  $z'-i'$ colors and spectral slopes over 
the $g', r'$, and $i'$ reflectance values, that were then used to assign 
to each asteroid a likely spectral type.  Since the authors were interested
in dealing with a complete sample, they limited their analysis to 
asteroids with absolute magnitude $H$ higher than 15.3, which roughly 
corresponds to asteroids with diameters larger than 5~km, and for which the
SDSS-MOC4 data-set is supposed to be complete. Since Carruba et al. (2015)
showed that the vast majority of asteroids that could survive 3.8 Gyr
of dynamical evolution in the post-LHB scenario should have diameters
larger than 5 km, we believe that this choice of limit in $H$ could
also be suitable for this research. For completeness, we also included 
asteroids with $H < 12.00$ with known spectral types from the Planetary Data 
System (Neese 2010) that are not part of SDSS-MOC4.  These objects
were assigned values of the $gri$ slope and $z' -i'$ colors at 
the center of the range of those of each given class.  We refer the 
reader to DeMeo and Carry (2013) and 
Carruba et al. (2014a) for more details on the procedure to 
obtain $gri$ slope and $z' -i'$ colors, and on how to assign asteroids
spectral classes based on these data.

Once a complete set of data with proper elements, obtained from the AstDys
site (http://hamilton.dm.unipi.it/cgi-bin/astdys/astibo, 
accessed on April $4^{th}$, 2015 (Kne\v{z}evi\'{c} and Milani 2003),
taxonomical and SDSS-MOC4 data on $gri$ slope and $z' -i'$ colors has been
computed for asteroids in each given asteroid family region, family 
membership is obtained using the Hierarchical Clustering Method (HCM) 
of Bendjoya and Zappal\'{a} (2002) in an extended domain of proper
elements and $gri$ slope and $z' -i'$ colors, where distances were computed
using the new distance metric:

\begin{equation}
d_{md} = \sqrt{d^2+C_{SPV}[(f_{gri} \cdot \Delta {gri})^2+ (\Delta (i-z))^2]},
\label{eq: carruba_metr}
\end{equation}

\noindent
where $d$ is the standard distance metrics in proper element domain 
defined in Zappal\'{a} et al. (1995) as:

\begin{equation}
d = na \sqrt{k_1 (\frac{\Delta a}{a})^2 +k_2(\Delta e)^2+k_3(\Delta \sin{(i)})^2}, 
\label{eq: stand_metr}
\end{equation}

\noindent 
where $n$ is the asteroid mean motion; $\Delta x$ the difference in proper
$a, e,$ and $\sin{(i)}$; and $k_1, k_2, k_3$ are weighting factors, defined as
$k_1$ = 5/4, $k_2$ = 2, $k_3$ = 2 in Zappal\'{a} et al. (1990, 1995).  
$\Delta {gri}$ and $\Delta {(i-z)}$ are difference between two neighboring 
asteroids $gri$ slopes and $z' -i'$ colors, $C_{SPV}$ is a constant equal to
$10^6$ (see also Carruba et al. 2013 for a discussion on the choice of this
constant) and $f_{gri} =0.027$ is a normalization factor to account for 
differences among the mean values of differences in $gri$ slope and $z' -i'$ 
colors (different values of $f_{gri}$ in the range 0.05-0.001 have been 
used without substantially affecting the output of the method).
We assigned families to their regions, defined as in Bro\v{z} et al. (2013) 
as the central main belt, the pristine region, the outer main belt and the 
outer highly inclined region, and compute proper elements and $gri$ slope and 
$z' -i'$ colors for all asteroids in each given region.  
We then computed nominal distance cutoff and stalactite diagrams with the
standard techniques described in Carruba (2010b), for all the seven Bro\v{z} 
families, and then assigned values of diameters and geometric albedos from the 
WISE catalog (Wright et al. 2010) to family members for which this information 
is available (other objects were assigned the values of geometric albedo of the
largest object in the family, and diameters computed using equation
1 in Carruba et al. 2003).  Finally the method of the Yarkovsky isolines
(see also Carruba et al. 2013) is applied to obtain an estimate of the maximum 
possible age of each given family.  In the next subsection we will 
discuss in detail the method for families in the outer main belt, 
results are similar for other regions.

\subsection{The outer main belt: the Themis, Meliboea, and Ursula families}
\label{sec: fam_outer}

The outer main belt is defined in Nesvorn\'{y} et al. (2015) as the region
between the 7J:-3A and 2J:-1A mean-motion resonances and $\sin{(i)} < 0.3$.
Three possible old C-complex families have been proposed in this region: 
the Themis, Meliboea, and Ursula groups.  We identified 5472 asteroids with
either taxonomical or SDSS-MOC4 information in the area, 3524 of which 
with data in the WISE data-set, and 2896 (82.2\%) with $p_V < 0.15$.
The value of the minimal distance cutoff $d_0$ was of 
157.5~m/s and the minimal number of objects to have a group statistically 
significant was 25.

\begin{figure*}

  \centering
  \begin{minipage}[c]{0.45\textwidth}
    \centering \includegraphics[width=3.in]{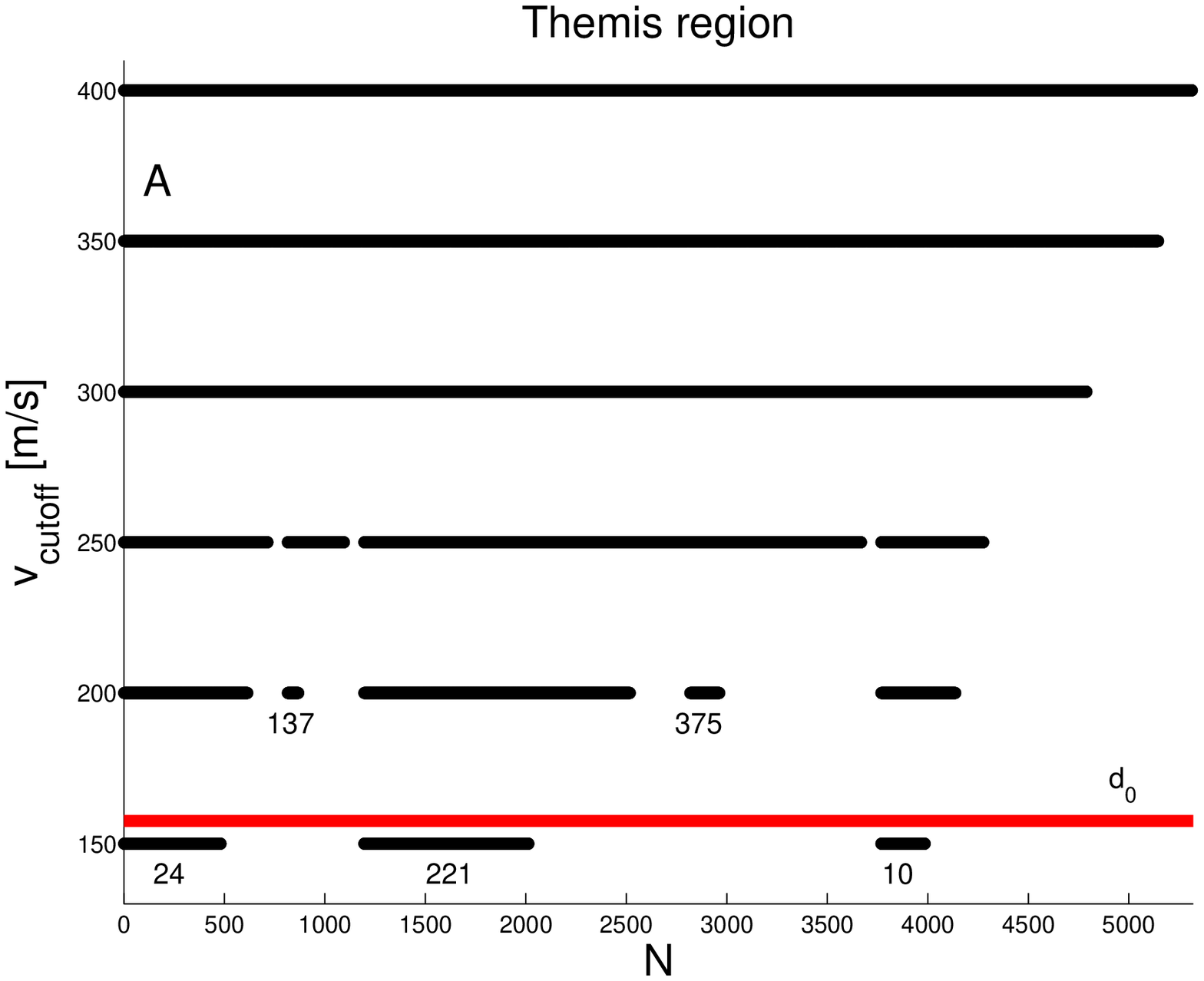}
  \end{minipage}%
  \begin{minipage}[c]{0.45\textwidth}
    \centering \includegraphics[width=3.in]{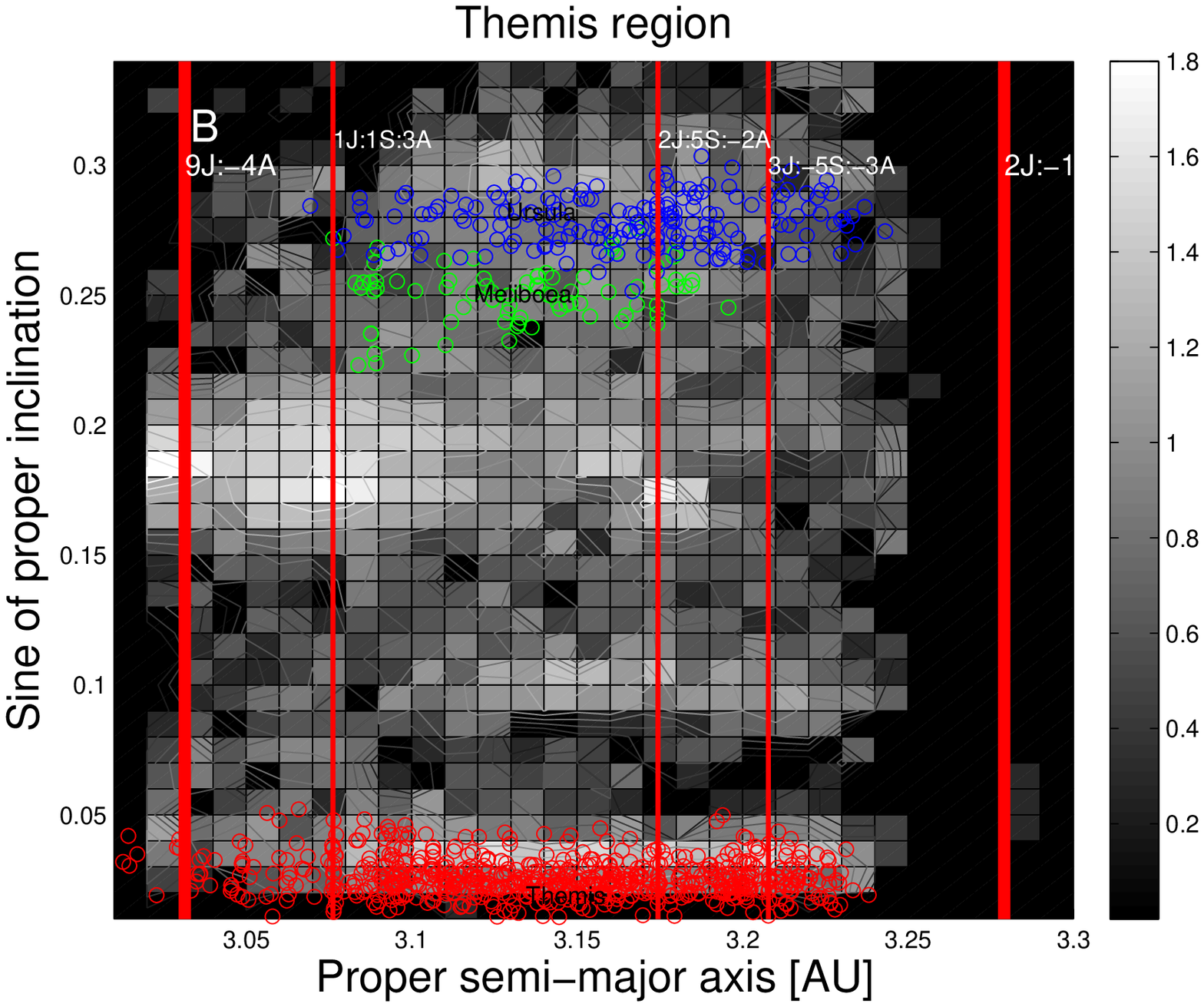}
  \end{minipage}

\caption{Panel A: stalactite diagram for the 5472 asteroids in the 
Themis region.  Asteroids belonging to a family are identified by a 
black dots. Panel B: contour plot of number density of asteroids
in the $(a,\sin{(i)})$ plane.   Red, green and blue circles show the 
location of members of the Themis, Meliboea, and Ursula families, 
respectively.}
\label{fig: meliboea}
\end{figure*}

Fig.~\ref{fig: meliboea}, panel A, displays a stalactite diagram for 
this region.  The families of Themis, Meliboea, and Ursula were all
visible at the nominal distance cutoff $d_0$, but not for lower values
of $d$.  137 Meliboea merges with the local background at $d = 300$~m/s. 
The family of 375 Ursula merges with Eos at cutoff of 250 m/s,
and with other minor groups at $d = 225$~m/s, so we choose to work 
with a cutoff of 220 m/s.  Other important families 
in the region were those of 221 Eos and 10 Hygiea.  Panel B of 
Fig.~\ref{fig: meliboea} displays a contour plot of number density 
of asteroids in the $(a,\sin{(i)})$ plane.   Whither tones are associated
with higher values of local number density. For our grid in this domain, 
we used 30 steps of 0.01 AU in $a$, starting from $a = 3.0~$AU,
and 34 steps of 0.01 in $\sin{(i)}$, starting from $\sin{(i)} = 0.0$. The
members of the identified Themis, Meliboea and Ursula families, after
taxonomical interlopers were removed (there were none in these cases), 
are shown as red, green, and blue circles, respectively.

\begin{figure}
  \centering
  \centering \includegraphics [width=0.45\textwidth]{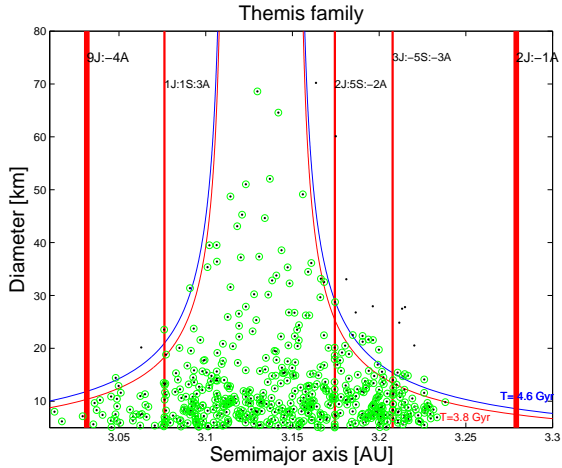}

\caption{Yarkovsky isolines age estimates for the Themis family.  Black 
dots identify members of the dynamical group without the taxonomical 
interlopers, and green circles show the position of members that are not
dynamical interlopers.  Vertical red lines display the location of 
mean-motion resonances, the red and blue lines displays isolines
of displacement from the family barycenter due to Yarkovsky effect
and close encounters with massive asteroids over 3.8 and 4.6 Gyr,
respectively.}
\label{fig: iso_themis}
\end{figure}

Finally, we also apply the method of Yarkovsky isolines (Carruba et al. 
2013) to the Themis, Meliboea, and Ursula families, using standard values of 
the parameters of this force for C-type groups from Bro\v{z} et al. (2013), 
Table 2.  Results are shown in Fig.~\ref{fig: iso_themis} for the Themis 
family.  The red and blue lines are the expected displacement of family
members over 3.8 and 4.6 Gyr because of the Yarkovsky effect and close
encounters with massive asteroids, assumed equal to 0.01~AU for
simplicity (Carruba et al. 2014a).  All asteroids were assumed to be 
initially at the family barycenter. This method yields maximum ages of 4.6 
Gyr for the three Themis, Meliboea, and Ursula families. 
Information on number of confirmed members and interlopers is given in 
Table~\ref{table: yarko_iso}, among other quantities.

\begin{table*}
\begin{center}
\caption{{\bf Orbital region, spectral complexes, number of asteroids in 
the orbital region, distance cutoff $d_0$, number of taxonomical interlopers, 
of dynamical interlopers, confirmed members, and 
maximum age estimates, inferred by Yarkovsky isolines, of the ``Bro\v{z}'' seven families.}}
\label{table: yarko_iso}
\vspace{0.5cm}
\begin{tabular}{|c|c|c|c|c|c|c|c|c|}
\hline
         &        &         &          &         &               &               &                   &   \\
Family   & Orbital& Number of& $d_0$   & Spectral& Number of tax.& Number of dyn.& Number of         & Maximum age \\
name     & region & asteroids& [m/s]   & Complex & interlopers   & interlopers   & confirmed members & estimate [Gyr]\\
         &        &         &          &        &                &               &                   &               \\
\hline
         &               &      &    &    &    &    &      &     \\
Eunomia  & Central mb    & 1416 & 155& S  & 33 &  1 & 1101 & 3.8 \\
Maria    & Central mb    & 1416 & 155& S  & 16 &  1 &  386 & 3.8 \\
Koronis  & Pristine zone & 1015 & 135& S  & 18 &  1 &  502 & 4.6 \\
Themis   & Outer mb      & 5472 & 220& CX & 0  & 10 &  642 & 4.6 \\
Meliboea & Outer mb      & 5472 & 220& CX & 0  &  4 &   78 & 4.6 \\
Ursula   & Outer mb      & 5472 & 220& CX & 0  & 11 &  172 & 4.6 \\
Alauda   & HI Outer mb   &  286 & 250& CX & 1  & 23 &  122 & 4.6 \\
         &               &      &    &    &    &    &      &     \\
\hline
\end{tabular}
\end{center}
\end{table*}

\section{Chronology}
\label{sec: chron}

Monte Carlo methods to obtain estimates of the family age and ejection 
velocity parameters were pioneered by Vokrouhlick\'{y} et al. (2006a, b, c) 
for the Eos and other asteroid groups.   They were recently modified
to account for the ``stochastic'' version of the YORP effect (Bottke et al. 
2015), and for changes in the past values of Solar luminosity 
(Vokrouhlick\'{y} et al. 2006a) for a study of dynamical groups in the 
Cybele region (Carruba et al. 2015).  Age estimates obtained including
the stochastic YORP effect in Carruba et al. (2015) were i) of better 
quality with respect those
obtained with the static version of YORP for old families, in terms of 
confidence level, and ii) tend to produce
younger age estimates.  We refer the reader to Bottke et al. 
(2015) and Carruba et al. (2015) for a more in depth description of 
the method.  Essentially, the semi-major
axis distribution of various fictitious families is evolved under
the influence of the Yarkovsky, both diurnal and seasonal versions, 
and YORP effect (and occasionally other effects such as close encounters 
with massive asteroids (Carruba et al. 2014a) or changes in past solar 
luminosity values (Carruba et al. 2015).  This method, however, ignores
the effect of planetary perturbations.  The newly obtained distributions 
of a $C$-target function computed with the relationship:

\begin{equation}
0.2H=log_{10}(\Delta a/C),
\label{eq: target_funct_C}
\end{equation}  

\noindent are then compared to the current $C$ distribution of real 
family members using a ${\chi}^2$-like variable ${\psi}_{\Delta C}$  
(Vokrouhlick\'{y} et al. 2006a, b, c), whose 
minimum value is associated with the best-fitted solution. 

\begin{figure}
  \centering
  \centering \includegraphics [width=0.45\textwidth]{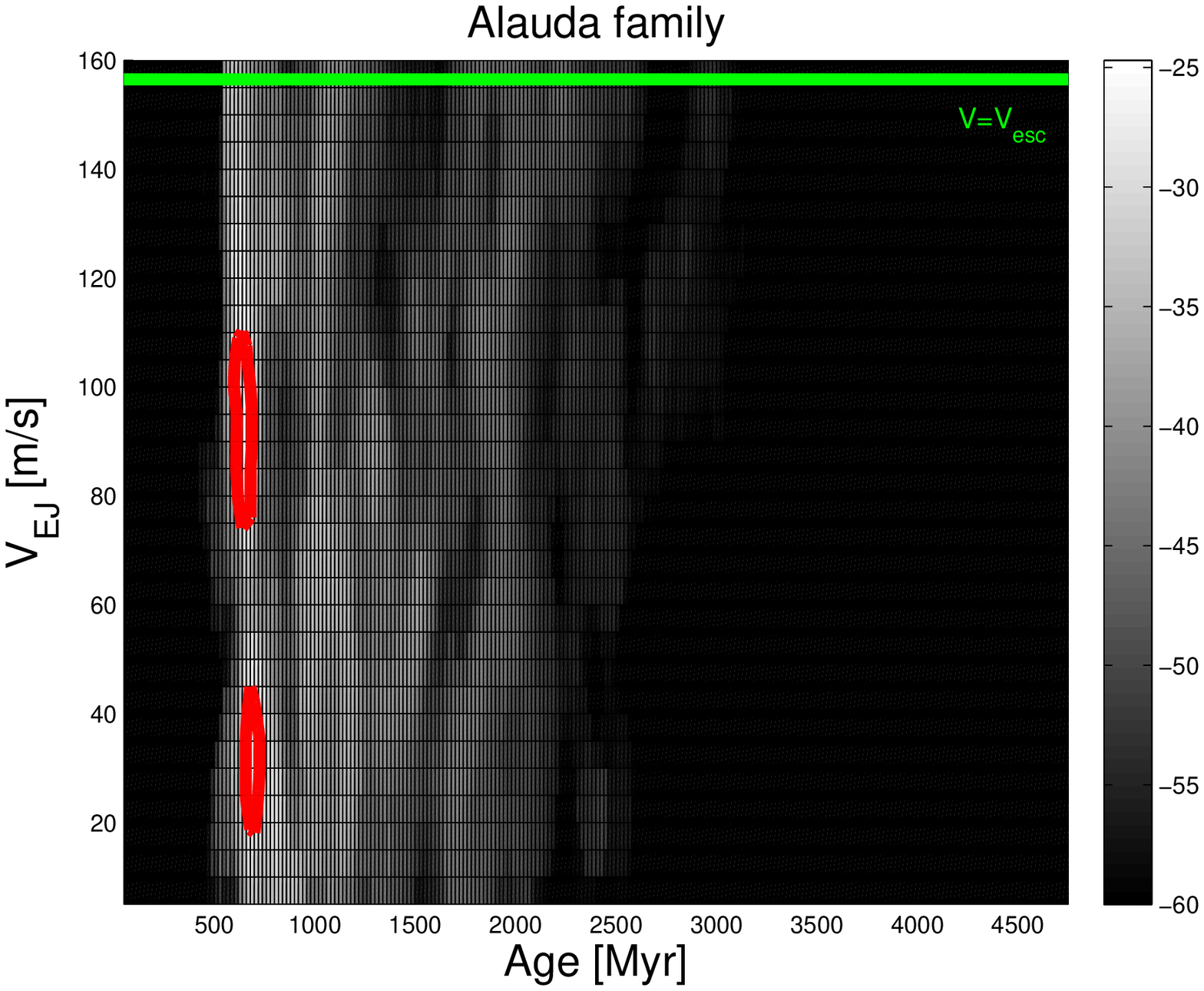}

\caption{Target function ${\psi}_{\Delta C}$ values in the ($Age,V_{EJ}$) plane 
for Alauda family. The horizontal green line display the value of 
the estimated escape velocity from the parent body.  The red
lines display the contour level of ${\psi}_{\Delta C}$ associated
with a 1-sigma probability that the simulated and 
real distribution were compatible.}
\label{fig: cont_alauda}
\end{figure}

We applied this method to six out of the seven ``Bro\v{z}'' families.
For the case of the Maria family, since this group was significantly
depleted at lower semi-major axis by interaction with the 3J:-1A 
mean-motion resonance, the number of remaining intervals in the $C$-target
function was too small to allow for a precise determination of the 
family age.  Fig.~\ref{fig: cont_alauda} displays ${\psi}_{\Delta C}$ values 
in the ($Age,V_{EJ}$) plane for Alauda family.  Values of the $V_{EJ}$ parameter,
that describes the spread in the terminal ejection velocities 
(Vokrouhlick\'{y} et al. 2006), tend to be lower than the estimated escape 
velocity from the parent body (Bottke et al. 2015), 156.5 m/s for the case 
of the Alauda family.  To estimate nominal values of the 
uncertainties associated with our estimate of the age and the $V_{EJ}$ 
parameter, here we used the approach first described in Vokrouhlick\'{y} et 
al. (2006a,b,c). First, we computed the number of degrees of freedom of the 
${\chi}^2$-like variable, given by the number of intervals in the $C$ 
distribution with more than 10 asteroids (we require a minimum number of 10 
asteroid per $C$ interval so as to avoid the problems associated with dividing 
by small number when computing ${\psi}_{\Delta C}$), minus 2, the number
of parameters estimated from the distribution.  Then, 
assuming that the ${\psi}_{\Delta C}$ probability distribution is given
by a incomplete gamma function of arguments ${\psi}_{\Delta C}$ and 
the number of degrees of freedom, we computed the value of ${\psi}_{\Delta C}$ 
associated with a 1-sigma probability (or 68.3\%) that the simulated and 
real distribution were compatible (Press et al. 2001).\footnote{We 
should caution the reader that other methods to estimate the uncertainties 
for the estimated parameters are also used in the literature.  For instance, 
one can compute the 1-, 2- or 3-sigma ${\chi}^2$ values and sum these to the 
minimum observed value of ${\chi}^2$ to obtain estimates of the errors at 
1-, 2-, or 3-sigma levels (Press et al. 2001).  Since age estimates for 
very old families, such as those studied in this work, tend to produce 
shallow minima, here we prefer to use the Vokrouhlick\'{y} et al. (2006a,b,c) 
approach, so as to provide a more limited range of estimated values. But 
larger values of the error estimates are possible.}

For the case of the Alauda family, we had 32 $C$ intervals with more
than 10 asteroids, and 30 degrees of freedom.  The Alauda family 
should be $640\pm50$~Myr old, with 
$V_{EJ} = 95^{+15}_{-20}$~m/s, and a secondary minimum at lower values of 
$V_{EJ}$.  Our results for the six families are summarized in 
Table~\ref{table: Chron_families}, were 
we display the name of the family, the estimated age, the values of 
$V_{EJ}$, and the limit used for ${\psi}_{\Delta C}$. 

\begin{table}
\begin{center}
\caption{{\bf Estimated age and ejection velocity parameter from 
Monte Carlo chronology for six ``Bro\v{z}'' families.}}
\label{table: Chron_families}
\vspace{0.5cm}
\begin{tabular}{|c|c|c|c|}
\hline
     &       &         &                   \\
Name &  Age  & $V_{EJ}$ & ${\psi}_{\Delta C}$ \\
     & [Myr] &  [m/s]  &                   \\
     &       &         &                   \\
\hline
     &       &         &                   \\
Eunomia  &$2020^{+650}_{-400}$  & $90^{50}_{-90}$ &18.24\\
Koronis  &$2360^{+2240}_{-2090}$& $70^{+60}_{-70}$ &12.59\\
Themis   &$1500^{+2650}_{-1010}$& $70^{+75}_{-70}$ &15.41\\
Meliboea &$640\pm10$         & $15^{+10}_{-5}$ & 7.02\\
Ursula   &$1060^{+3540}_{1060}$ & $45\pm45$     & 4.30\\
Alauda   &$640\pm50$         & $95^{+15}_{-20}$&27.77\\
     &       &         &                         \\     
\hline
\end{tabular}
\end{center}
\end{table}

Overall, we found that no family is nominally older than 2.7 Gyr, 
but uncertainties are too large for the Koronis, Themis, and Ursula
families for a positive conclusion to be reached.  We compared
our results with the maximum age estimates from Bro\v{z} et al. (2013)
and from the more recent work of Spoto et al. (2015), that found
estimates for the family ages with a V-shape method in the $(a,1/D)$ domain.  
Table~\ref{table: max_ages} shows the maximum estimated ages from this work
(mean values plus errors),  from Bro\v{z} et al. (2013) and from Spoto 
et al. (2015)\footnote{In Spoto et al. (2015) the authors found estimates 
for the left and right semi-major axis distributions of family members with 
respect to the family barycenter.  Results in Table~\ref{table: max_ages} are 
for the maximum possible estimates, among the two, when available (for 
the Maria and Ursula there was data for just one of the family wings, there
was no estimate available for the Meliboea and Alauda families.  
The Eunomia parent body may have experienced two or more impacts, according
to these authors. Table~\ref{table: max_ages} reports the estimated
age of the oldest impact).}.
 
\begin{table}
\begin{center}
\caption{{\bf Estimated maximum ages from this work, Bro\v{z} et al.
(2013), and Spoto et al. (2015).}}
\label{table: max_ages}
\vspace{0.5cm}
\begin{tabular}{|c|c|c|c|}
\hline
     &            &              &            \\
Name & Current    &  Bro\v{z} et & Spoto et   \\
     & estimates  &  al. (2013)  & al. (2015) \\
     &   [Gyr]    &    [Gyr]     &  [Gyr]     \\
     &            &              &            \\
\hline
     &            &      &      \\
Eunomia  & 2.67   & 3.00 & 2.35 \\
Maria    &        & 4.00 & 2.35 \\
Koronis  & 4.60   & 3.50 & 2.24 \\
Themis   & 4.15   & 3.50 & $<$4.60\\
Meliboea & 0.65   & $<$3.00&      \\
Ursula   & 4.60   & $<$3.50& 4.52 \\
Alauda   & 0.69  & $<$3.50&      \\
     &            &      &       \\
\hline
\end{tabular}
\end{center}
\end{table}

To within the uncertainties, our estimates tend to be in agreement 
with previous results, with the two possible exceptions of the Alauda 
and Meliboea families, that could potentially be younger than previously
thought (but uncertainties for these two families may be larger if different
approaches for the error on the ${\chi}^2$-like variable were used.)
Masiero et al. (2012) investigated the effect that changes in the nominal
values of the parameters affecting the Yarkovsky and YORP forces may have on the
estimate of the age of the Baptistina family and found that the 
parameters whose values most affected the strength of the Yarkovsky 
force were the asteroid density and the thermal conductivity.  
Since the largest variations were observed for changes in the values of 
the family thermal conductivity, for the sake of brevity in this work 
we concentrate our analysis on this parameter.
Maximal ages can be found if one consider a value of the thermal conductivity
of $K= 0.1$~W/m/K.  We repeated our analysis for the six ``Bro\v{z}'' 
families, and this larger value of $K$, and our results are summarized in 
Table~\ref{table: Chron_families_K}, where we display the estimated ages, 
ejection velocity parameter, and maximum possible age.

\begin{table}
\begin{center}
\caption{{\bf Estimated age, ejection velocity parameter, and 
maximum possible age for six ``Bro\v{z}'' families, when 
$K= 0.1$~W/m/K.}}
\label{table: Chron_families_K}
\vspace{0.5cm}
\begin{tabular}{|c|c|c|c|}
\hline
     &       &         &           \\
Name &  Age  & $V_{EJ}$ & Max. Age  \\
     & [Myr] &  [m/s]  &  [Gyr]    \\
     &       &         &           \\
\hline
     &       &         &           \\
Eunomia  &$>4600$          & $80\pm{40}$   &$>$ 4.6\\
Koronis  &$4500^{+100}_{-300}$& $5^{+125}_{-5}$     &4.6\\
Themis   &$2240^{+2360}_{-1420}$& $60^{+80}_{-60}$  &4.6\\
Meliboea &$1250\pm50$      & $10^{+20}_{-10}$    &1.3\\
Ursula   &$3150^{+1450}_{-3150}$ & $15^{+80}_{-15}$ &4.6\\
Alauda   &$1020^{+80}_{-40}$    & $95^{+20}_{-10}$ &1.1\\
     &       &         &           \\     
\hline
\end{tabular}
\end{center}
\end{table}

The ages of the S-type families Koronis and Eunomia in these simulations
was larger then 4.5 Gyr, but that it is just an artifact caused by the 
improbably high value of $K$ (0.1~W/m/K) used for these simulations 
(typical values of $K$ for S-type families are of the order of 0.001~W/m/K).  
More interesting were the results for the C-complex groups:
while the maximum possible ages for the Themis and Ursula families were 
beyond 3.0 Gyr, none of the groups, even in this very favorable scenario,
has nominal ages old enough to reach the earliest estimates for the LHB 
(3.8 Gyr ago).  The implications of this analysis will be further 
explored in the next section.

\section{Dynamical evolution of old families}
\label{sec: dyn_old}

To study the possible survival of any of the largest members of the oldest
main belt family over $\simeq$ 4 Gyr, we performed simulations with 
the $SYSYCE$ integrator (Swift$+$Yarkovsky$+$Stochastic
YORP$+$Close encounters) of Carruba et al. (2015), modified to also
account for past changes in the values of the solar luminosity.
The numerical set-up of our simulations was similar to what discussed
in Carruba et al. (2015): we used the optimal values of the Yarkovsky 
parameters discussed in Bro\v{z} et al. (2013) for C- and S-type asteroids,
the initial spin obliquity was random, and normal reorientation timescales 
due to possible collisions as described in Bro\v{z} (1999) were 
considered for all runs. We integrated our test particles  
under the influence of all planets, and obtained synthetic proper elements
with the approach described in Carruba (2010b).

We generated fictitious families with the ejection parameter $V_{EJ}$
found in Sect.~\ref{sec: chron} (for the Maria family we used the same 
value found for the Eunomia group, i.e., $V_{EJ}= 90~m/s$), and integrated 
these groups over 4.0 Gyr. Also, since only bodies larger than 4 Km in diameter
were shown to survive in the Cybele region over 4.0 Gyr, following
the approach of Carruba et al. (2015) we generated families
with size-frequency distributions (SFD) with an exponent $-\alpha$ that 
best-fitted the cumulative distribution equal to 3.6, a fairly typical
value, and with diameters in the range from 2.0 to 12.0 km.
The number of simulated objects was equal to the currently observed
number of family members with diameters between 2.0 and 12.0~km.

For each of the simulated families, we computed the fraction of objects 
that remained in a box defined by the maximum and minimum value of 
$(a,e,\sin{(i)})$ associated with current members of the seven families 
as a function of time, the fraction of objects with $4< D < 6$ km,
and the time evolution of the $-\alpha$ exponent.  These parameters
will help estimating the dynamical evolution of the simulated families:
the lower the values of these numbers, the more evolved and diffused 
should be the family.  Also, to quantify
the dispersion of the family members as a function of time, we also computed
the nominal distance velocity cutoff $d_0$ for which two nearby asteroids are 
considered to be related using the approach of Beaug\'{e} and Roig (2001),
that defines this quantity as the average minimum distance between all
possible asteroid pairs, as a function of time (typical values are of
the order of 50 m/s, significantly larger values would indicate that the
family was dynamically dispersed beyond recognition). 

\begin{figure*}
  \centering
  \centering \includegraphics [width=0.65\textwidth]{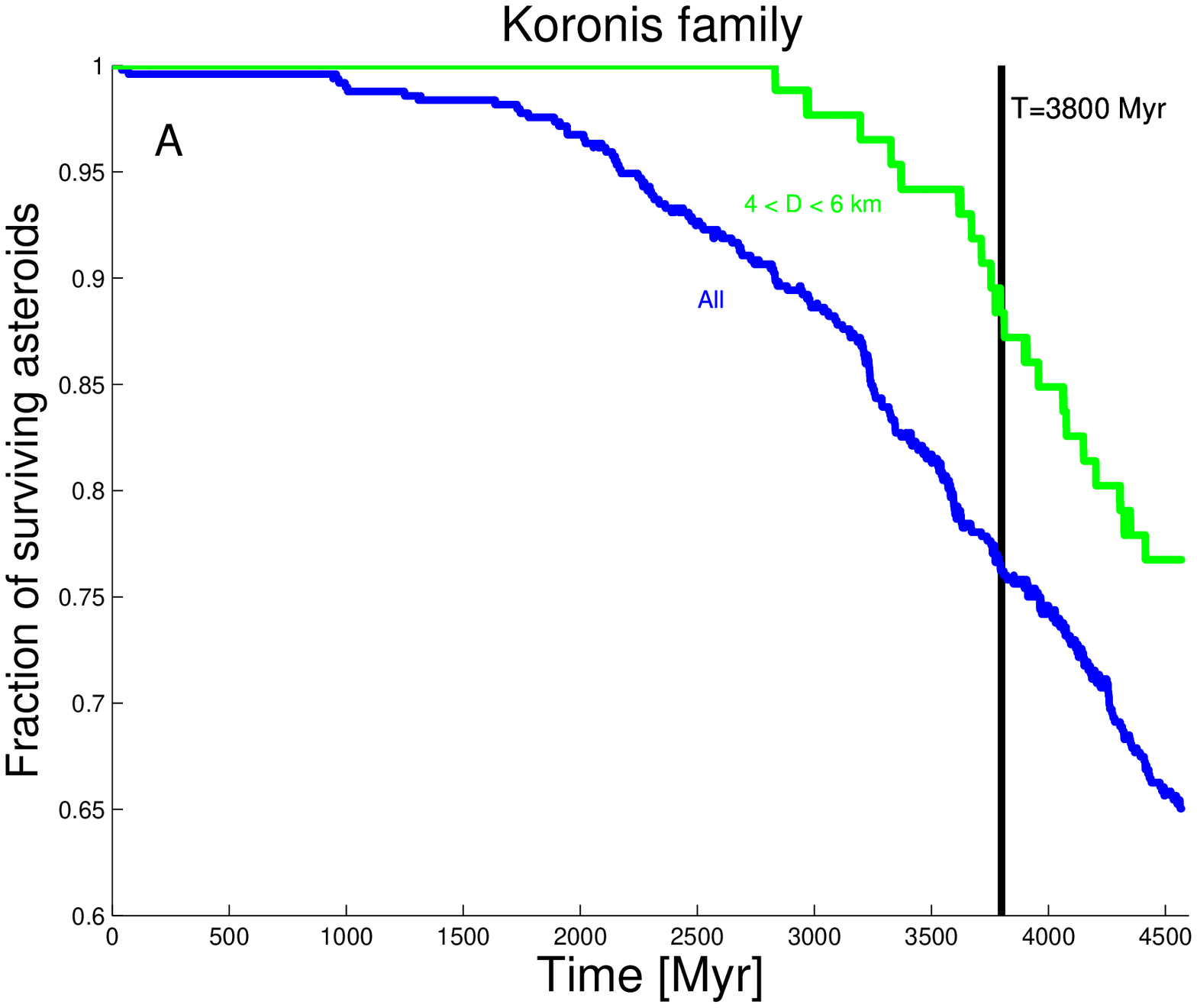}

  \centering
  \begin{minipage}[c]{0.45\textwidth}
    \centering \includegraphics[width=3.in]{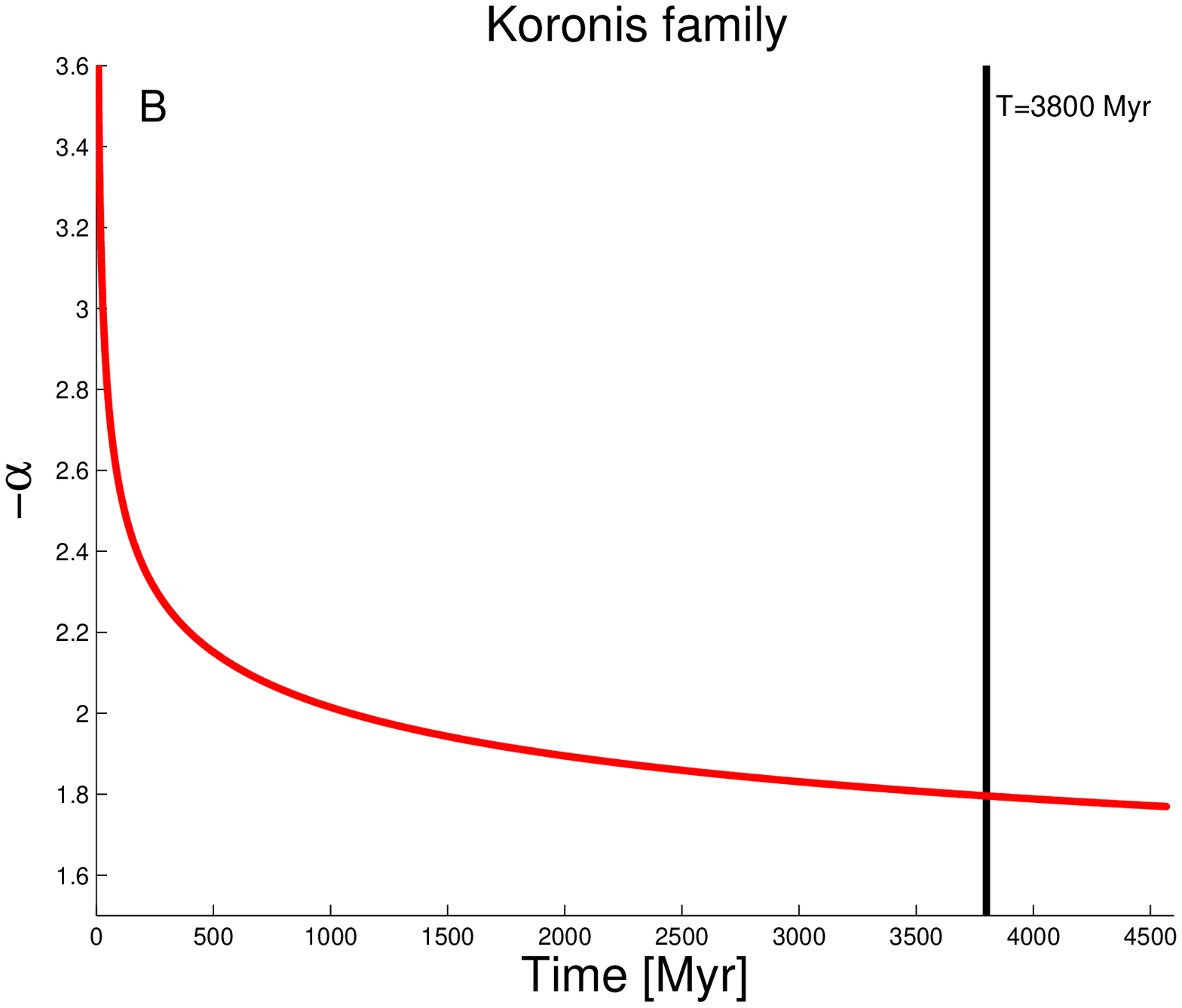}
  \end{minipage}%
  \begin{minipage}[c]{0.45\textwidth}
    \centering \includegraphics[width=3.in]{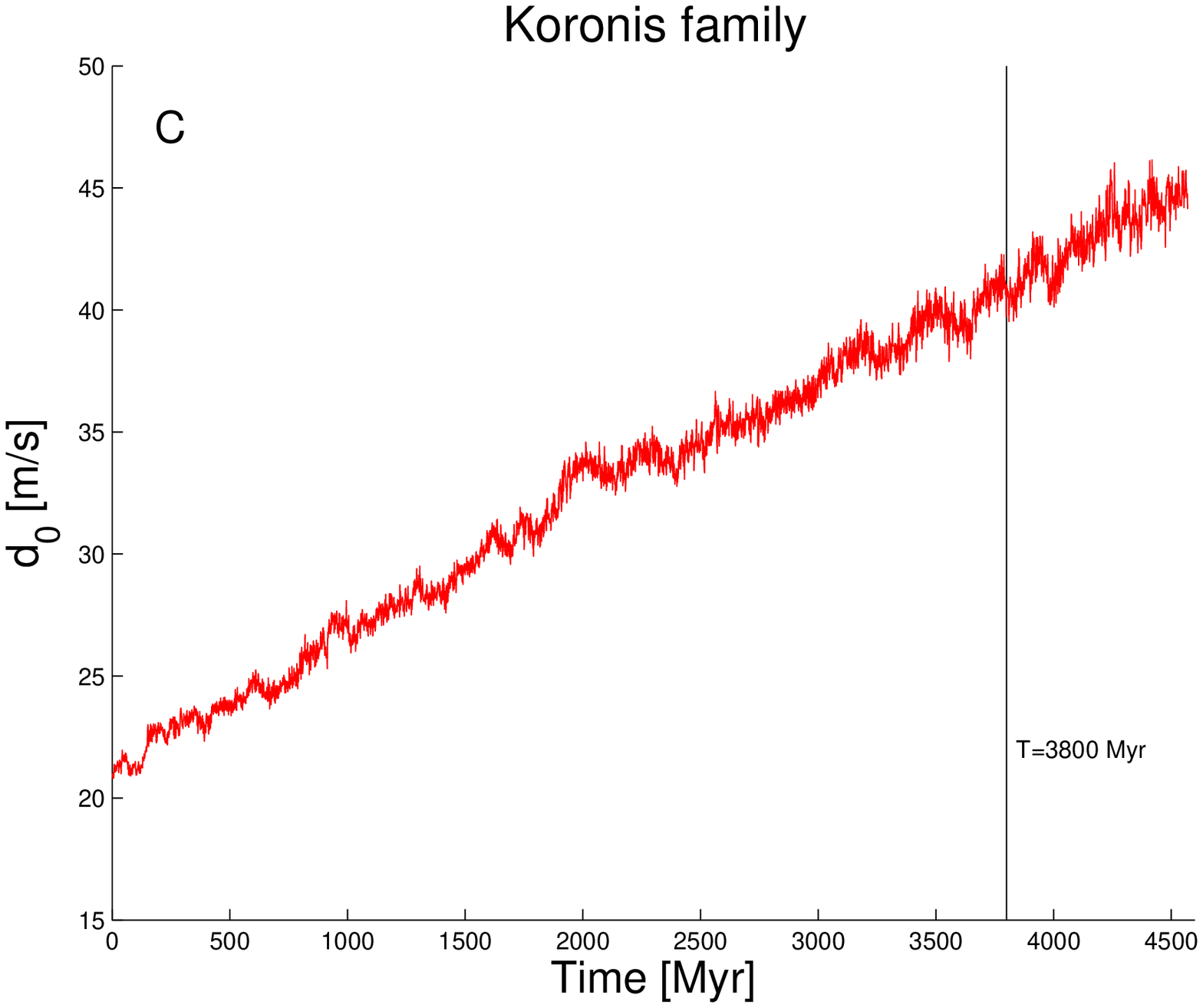}
  \end{minipage}

\caption{Panel A: the fraction of particles of all sizes (blue line) 
and with $4 < D < 6$ km (green line) that remained in the Koronis orbital
region as a function of time.  Panel B:  the time evolution of the 
$-\alpha$ exponent of the size cumulative distribution.  Panel C:
time evolution of $d_0$, the nominal distance velocity cutoff necessary
to recognize two objects as members of the dynamical family.}
\label{fig: kor_sysyce}
\end{figure*}

Our results for the Koronis family are shown in Fig.~\ref{fig: kor_sysyce},
where we display in panel A the time evolution of the fraction of all asteroids 
remaining in the Koronis family region (blue line) and of the objects
with $4< D < 6$ km (green line).  Panel B shows the time evolution of the 
$-\alpha$ exponent, while panel C displays $d_0$ as a function of time.
The Koronis family is in a dynamically less active region, so a larger 
fraction of its original population survive the simulation, but this is
not the case for all investigated families.  Our results at $t = 3.8$~Gyr,
the minimum estimated age for the end of the late heavy bombardment,
are reported in Table \ref{table: threshold_time}.

\begin{table*}
\begin{center}
\caption{{\bf Value at $t = 3.8$~Gyr of the fraction of particles with all 
diameters surviving the simulation (second column), of those 
with$ 4< D < 6$ km (third column), of the $-\alpha$ exponent (fourth
column), and of $d_0$ (fifth column).}}
\label{table: threshold_time}
\vspace{0.5cm}
\begin{tabular}{|c|c|c|c|c|}
\hline
     &                       &                       &           &       \\
Name & Fraction of surviving & Fraction of surviving & $-\alpha$ & $d_0$ \\
     & asteroids (all sizes) [\%] & asteroids ($4< D < 6$ km) [\%]&     & [m/s] \\
     &                       &                       &           &       \\
\hline
         &        &       &      &      \\
Eunomia  & 36.8   & 53.3  &  1.7 & 77.3 \\ 
Maria    & 72.7   & 92.0  &  1.7 & 63.5 \\
Koronis  & 76.2   & 88.3  &  1.8 & 42.2 \\
Themis   & 69.5   & 89.4  &  2.0 & 92.3 \\
Meliboea &  9.0   & 16.6  &  2.4 &375.2 \\
Ursula   & 65.1   & 92.1  &  2.0 &137.4 \\
Alauda   & 22.1   &  8.5  &  1.9 &171.4 \\
     &            &        &            \\
\hline
\end{tabular}
\end{center}
\end{table*}

\begin{figure*}
  \centering
  \begin{minipage}[c]{0.45\textwidth}
    \centering \includegraphics[width=3.in]{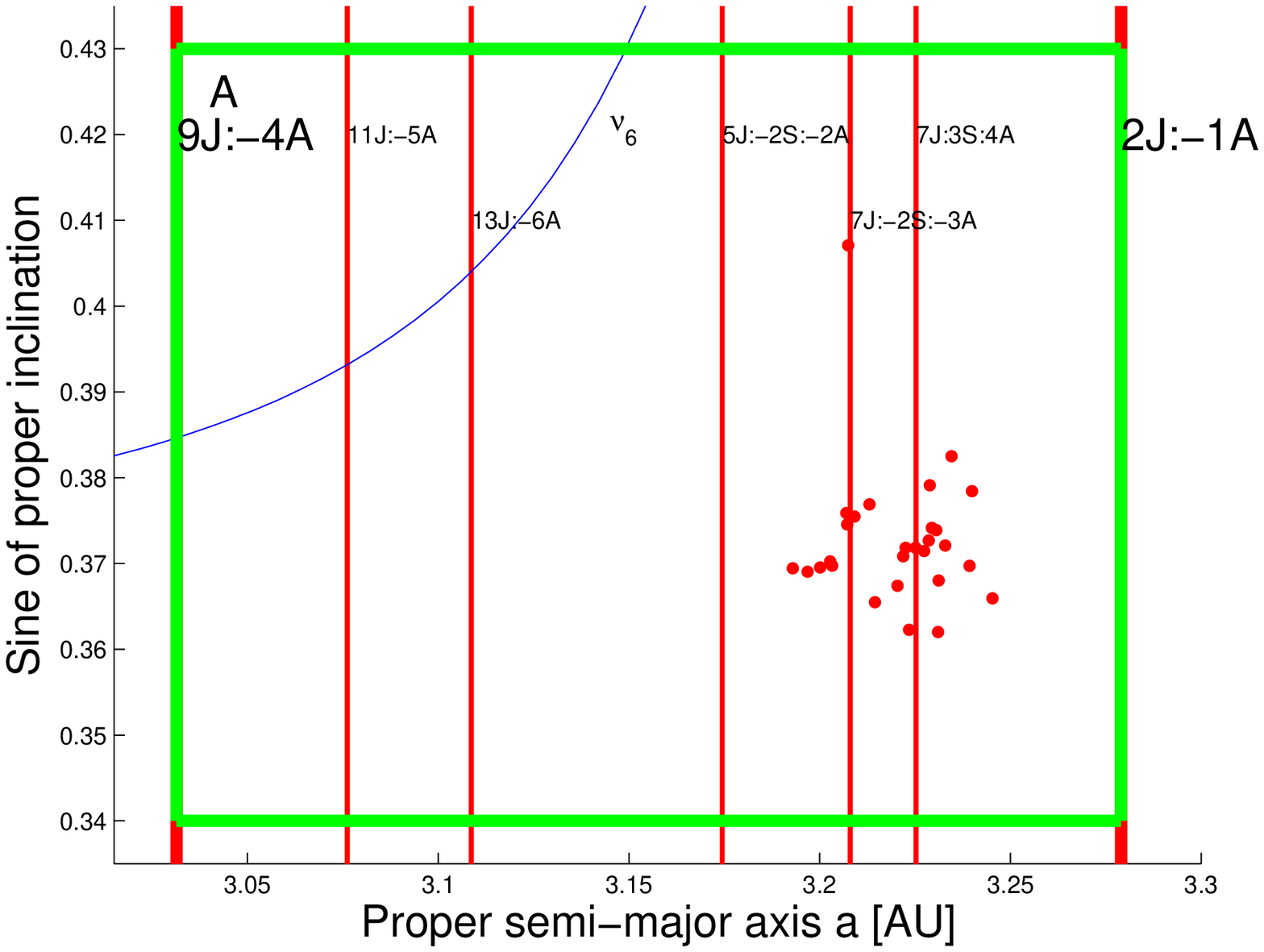}
  \end{minipage}%
  \begin{minipage}[c]{0.45\textwidth}
    \centering \includegraphics[width=3.in]{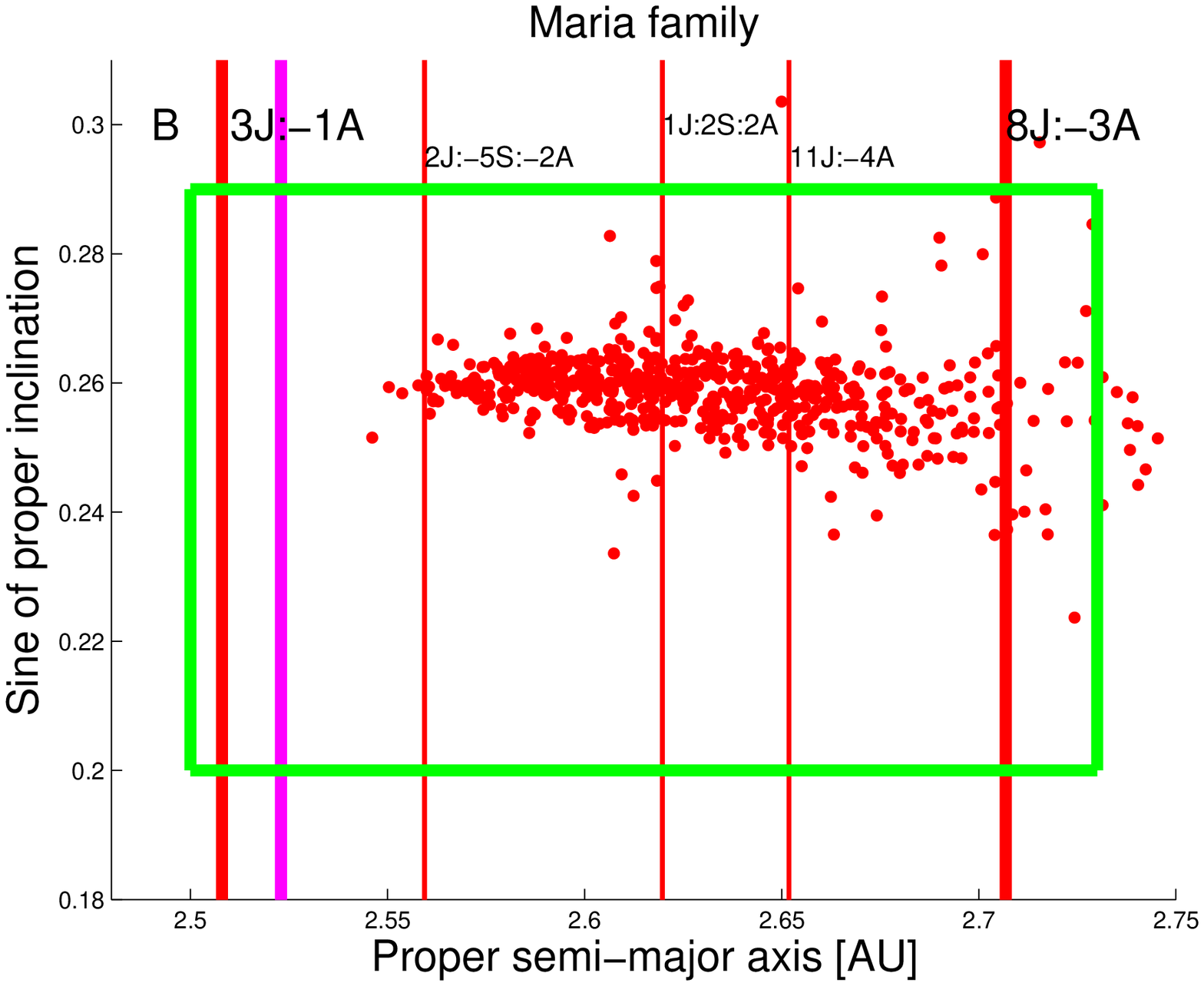}
  \end{minipage}

\caption{Projection at $t = 3.8$~Gyr in the $(a,\sin{(i)})$ plane of 
members (red full dots) of the simulated Alauda (panel A) and Maria (panel B) 
families.  Vertical red lines display the location of the main local 
mean-motion resonances, the magenta vertical line in panel A refers to the 
approximate location of the chaotic layer near the boundary of the 3J:-1A 
mean-motion resonance, where particle are unstable on time-scales of 100 Myr.  
The blue line in panel B show the approximate location of the ${\nu}_6$ secular
resonance.  Green lines define the boundary in the $(a,\sin{(i)})$ plane
of the orbital region of the two families.}
\label{fig: Scatter_alauda_maria}
\end{figure*}

Overall, the Meliboea and Alauda synthetic families were dispersed beyond
recognition.   All families had values of $-\alpha$ at $t = 3.8$~Gyr much
shallower than the initial value (3.6), and compatible with typical values
of background asteroids ($\simeq 2$).  This does not necessarily mean
that all paleo-families should be characterized by a shallow SFD.  
The initial SFD could have been much steeper, and collisionary 
evolution could have replenished the population of asteroids at smaller
sizes.  There are indeed some indications that some potential paleo-families,
such as Itha, could be characterized by a relatively steep SFD
(Bro\v{z} et al. 2012).  However, dynamical effects alone indeed tend to 
remove smaller size bodies and to produce families with shallower
SFD.  In regimes where dynamical effects are prodominant and the initial
SFD was not too steep, we would expect paleo-families to be characterized
by a shallow SFD.  Also, according to the values
of $d_0$ found in this work, only the synthetic Koronis, Maria, and possibly
Eunomia family would be recognizable, with some difficulties, with respect
to the background (typical values of $d_0$ depends on the local
density of asteroids, but are usually of the order of 50-60 m/s).  To 
help visualize the difference between a completely dispersed family,
such Alauda, and a relatively well preserved one, such Maria, we show
in Fig.~\ref{fig: Scatter_alauda_maria} a projection in the $(a,\sin{(i)})$
of the outcome of our simulations at $t = 3.8$~Gyr.  While only a handful
of the largest members of Alauda survived up to this time, the simulated 
Maria family, while dispersed, could still be recognizable in this
domain.

Can any paleo-family still be observable today?  As we discussed, paleo-families
would be difficult to recognize with traditional methods such as HCM, 
being characterized by a shallow SFD,
a significant depletion in small family members (those less than 5 km in 
diameter), and a large spread among the surviving members.  Paleo-families
belonging to fairly typical taxonomical classes, such as C- and S-type, 
would be extremely hard to recognize.  It was however proposed that V-type
asteroids in the Eunomia orbital region could have been fragments of a 
paleo-Eunomia family associated with the disruption of Eunomia parent body
crust (Carruba et al. 2007, 2014b).  We checked the $-\alpha$ value of 
the 16 V-type photometric candidates SFD currently in the Eunomia orbital 
region (defined according to our box criteria), and we found a value of 1.95.  
While this result should be considered with caution, given the limited
number of V-type asteroids in the region and possible limitations caused
by observational incompleteness, the very shallow SFD of these objects 
suggests, in our opinion, that an origin from a paleo-Eunomia family is not 
incompatible with the results of this work.

\section{Conclusions}
\label{sec: conc}

In this work we:

\begin{itemize}

\item Identified members of the seven old ``Bro\v{z}'' families in 
a new domain of proper elements, $gri$ slope and $z'-i'$ colors.
Once taxonomical and dynamical interlopers were removed, preliminary
estimates of the maximum possible family ages were obtained.

\item Used a ``Yarko-YORP'' Monte Carlo approach (Carruba et al. 2015)
that includes the effects of the stochastic YORP effect of Bottke et
al. (2015) and past changes in values of the solar luminosity to obtain
refined estimates of the family ages, when possible.  Our nominal age 
estimates are lower than results of other groups that did not consider
the stochastic YORP effect, as expected, but compatible to within the 
uncertainties.  Even allowing for the 
maximum possible value in the thermal conductivity of the simulated
families, no CX-complex group could have a nominal age dating 
from the latest phases of the LHB.

\item Simulated with the $SYSYCE$ symplectic integrators 
(Carruba et al. 2015) that accounts for the Yarkovsky and stochastic 
YORP effects, and past changes in solar luminosity, the dynamical 
evolution of members of fictitious original seven ``Bro\v{z}'' families.
Under the assumptions of our model (no collisional evolution, and an
initial SFD with a $-\alpha$ exponent for the population of objects
with $2 < D < 12$~km equal to 3.6), any ``paleo-family'' that formed 
between 2.7 and 3.8~Gyr ago would 
be characterized by a very shallow size-frequency distribution, a depletion
in smallest ($D < 5$ km) members, and a significant spread among the
surviving fragments.  Only families in dynamically less active regions,
such as the Koronis family in the pristine zone of the main belt, could have 
potentially partially survived 3.8~Gyr of dynamical evolution and not
be completely dynamically eroded.  The V-type asteroids in the 
Eunomia orbital region are characterized by a very shallow SFD, and could
potentially be compatible with a paleo-Eunomia family, as suggested
in the past (Carruba et al. 2007, 2014b).

\end{itemize}

Overall, the main result of this work is that some paleo-families,
particularly the initially most numerous ones, or those in dynamically less
active parts of the main belt, could still be visible today, but 
would be of rather difficult identification,
especially for the case of families belonging  to fairly typical taxonomical
types, such as the C- and S-types.  Other effects not considered in this
work, such as collisional cascading or comminution (Bro\v{z} et al. 2013),
close encounters with massive asteroids (Carruba et al. 2003), secular
dynamics involving massive asteroids (Novakovi\'{c} et al. 2015) could
all have contributed to further disperse paleo-family members, perhaps
beyond recognition.  Yet the quest for the identification of a paleo-family
remain, in our opinion, a very worthy subject of research in asteroid
dynamics.  If such family could be found, such as is possibly the case
for the V-type asteroids in the Eunomia orbital region, it could
provide precious clues about a very early stage of our Solar System.
Finding and identifying paleo-families remains therefore a very value line 
of research in asteroid dynamics.

\section*{Acknowledgments}
We thank the reviewer of this paper, Miroslav Bro\v{z}, for comments
and suggestions that improved the quality of this work.
We would like to thank the S\~{a}o Paulo State Science Foundation 
(FAPESP) that supported this work via the grants 14/06762-2 and 14/24071-7), 
and the Brazilian National Research Council (CNPq, grant 305453/2011-4).
DN acknowledges support from the NASA Solar System Working (SSW) program. 
The first author was a visiting scientist at the Southwest Research Institute
in Boulder, CO, USA, when this article was written.
This publication makes use of data products from the Wide-field 
Infrared Survey Explorer WISE and NEOWISE, which are a joint 
project of the University of California, Los Angeles, and the Jet 
Propulsion Laboratory/California Institute of Technology, funded by the 
National Aeronautics and Space Administration.

\bsp

\label{lastpage}

\end{document}